\documentclass[prd,nofootinbib,showpacs,twocolumn,floatfix]{revtex4}
\usepackage{latexsym}
\usepackage{dcolumn}
\RequirePackage{graphicx}

\newcommand{\MET}{{\slash\!\!\!\!E_T}}

\newcommand{\fbi}{\mathrm{fb}^{-1}}

\newcommand{\be}{\begin{equation}}
\newcommand{\ee}{\end{equation}}
\newcommand{\bea}{\begin{eqnarray}}
\newcommand{\eea}{\end{eqnarray}}

\begin{document}

\preprint{IIT-CAPP-10-03}

\title{Charm tagging and the $H \to W^+W^-\to l\nu c j$ semi-leptonic channel}

\author{Arjun~Menon}
\affiliation{Illinois Institute of Technology,
Chicago, Illinois 60616-3793, USA}
\author{Zack~Sullivan}
\email{Zack.Sullivan@IIT.edu}
\affiliation{Illinois Institute of Technology,
Chicago, Illinois 60616-3793, USA}

\date{June 5, 2010}

\begin{abstract}
  We introduce a method to discover the Higgs boson at the Large
  Hadron Collider (LHC) through its decay to $W^+W^-$, where one boson
  decays to leptons, and the other decays to $c+$jet.  This mechanism
  is complementary to the decay into dileptons, but has the potential
  to measure the invariant mass peak of the Higgs boson, and to avoid
  large recently discovered QCD backgrounds from heavy flavor decays.
  In addition, this mechanism motivates the study and creation of a
  dedicated charm jet tagger at LHC experiments.  Existing charm jet
  tagging, in the form of fakes to bottom jet tagging, provides
  sensitivity to a standard model Higgs boson that is comparable to
  $WW$ fusion.  A 50\% charm tagging efficiency in the relevant
  kinematic range could allow an independent $5\sigma$ discovery of a
  165 GeV Higgs boson in 7 $\fbi$ of integrated luminosity at a
  14~TeV machine, or exclusion with a 7~TeV collider.
\end{abstract}


\maketitle

\section{Introduction}
\label{sec:introduction}

The Higgs boson is the only particle of the standard model that has
not been discovered.  The ATLAS and CMS experiments at the Large
Hadron Collider (LHC) search for a standard model-like Higgs boson
through its decay to pairs of $W$ bosons for masses of $140 \mathrm{\
  GeV} \alt m_H \alt 200$~GeV where both $W$ bosons decay leptonically
\cite{Aad:2009wy,Ball:2007zza}.  While $H\to WW\to l^+l^-\nu\bar\nu$
has been the dominant search mode in this mass range for more than a
decade \cite{Barger:1990mn,Dittmar:1996ss,Han:1998ma}, there are two
challenges associated with the dilepton plus missing energy $\MET$
search.  First, the loss of two neutrinos means that a Higgs mass
cannot be directly reconstructed; instead it relies on a solid
understanding of the measured kinematic variables in the presence of
radiation effects and backgrounds \cite{Aad:2009wy,Ball:2007zza}.
Second, it has been shown that at higher instantaneous luminosities
there is a large and poorly determined background due to heavy-quark
decays into isolated leptons that can be eliminated through the use of
tighter cuts \cite{Sullivan:2006hb}.  In this paper we propose the use
of a complementary final state of Higgs boson decay, $H\to WW \to l\nu
c j$, which avoids both of these issues, and with improved charm
tagging, could have a sensitivity for discovery comparable to the
dilepton$+\MET$ search.

The semi-leptonic decay channel $H \to W^+ W^- \to l\nu jj$ has been
studied previously in the literature and was found to be not promising
for low masses due to the large $W jj$ background~\cite{Han:1998ma}.
For large masses (typically 300 GeV and above) Higgs boson studies
\cite{Aad:2009wy} predict reduced backgrounds, and $l\nu jj$ forms one
of the main channels for discovery.  Recent interest in this channel
at lower masses has been ignited by Ref.~\cite{Dobrescu:2009zf}, which
demonstrated the existence of angular correlations in the $l\nu jj$
channels between the plane of the jets and the leptons that may help
reduce the $Wjj$ background.  Nevertheless, the $Wjj$ background is
still significant, and makes a discovery in the $l\nu jj$ final state
difficult.  In this paper we identify a key ingredient is the use of
charm jet tagging to pick out the final state in which one of the jets
contains a charmed hadron.  With this addition, the $Wjj$ background
is greatly reduced.  Furthermore, by identifying the charm jet, a
unique assignment can be made to the angular correlations between all
four decay products --- allowing for much stronger cuts that enable a
clean extraction of the $l\nu c j$ signal, and a reconstruction of a
Higgs boson mass peak.

The essential ingredient to measuring Higgs boson production in the
semi-leptonic channel is the use of charm tagging.  Currently there
are no dedicated reconstruction algorithms for charm jet tagging in
the public ATLAS or CMS analysis codes.  However, charm jets are
already reconstructed as fakes to ``bottom jets'' at roughly a
10--15\% rate \cite{Carena:2000yx,PGS4}.  A second goal of this paper
is to advocate for the creation of a dedicated charm tagger with an
acceptance of closer to 50\%.  Beyond enabling the discovery of a
Higgs boson, a charm tagging capability is important for many other
processes.  As the first measurement of $Wc$ has shown \cite{:2007dm},
the ability to identify charm leads to important constraints on the
$s$ parton distribution function \cite{Tung:2006tb}.  In models of
supersymmetry, a light top squark can decay to charm via a
flavor-changing neutral current $\widetilde{t}\to c
\widetilde{\chi}_0$ \cite{SUSYT}.  In the so-called ``buried Higgs''
scenario it is possible for a Higgs boson to decay to four charm
quarks \cite{Bellazzini:2009xt}.  In this paper we consider a
continuum of scenarios: from the use of charms currently mistagged as
``$b$ jets'', through use a dedicated ``$c$-jet'' tagger with roughly
a 50\% acceptance.  We find in all cases that a 100\% acceptance of
events with of $b$ quarks faking $c$ jets does not change our results.

We describe the two key ingredients of our analysis, charm tagging and
angular correlations in Sec.\ \ref{sec:charmtag}.  In Sec.\
\ref{sec:sim}, we discuss a detailed list of cuts to reconstruct a
Higgs boson peak above background.  We then present the expected
signal significance at a 14 TeV LHC for a range of charm jet tagging
efficiencies, and contrast the results with the exclusion reach at a 7
TeV LHC.  We conclude with a summary of key points and future
directions.

\section{Charm tagging and angular correlations}
\label{sec:charmtag}

\begin{figure}[htb]
\centering
\includegraphics[width=0.4\columnwidth]{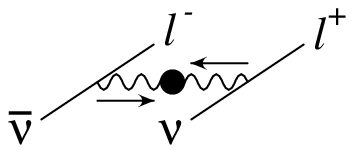}
\includegraphics[width=0.4\columnwidth]{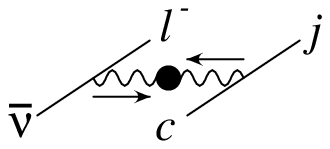}
\caption{Angular correlation between leptons in (left) $H\to W^-W^+\to
  l^-\bar\nu l^+\nu$ and (right) $H\to W^-W^+\to l^-\bar \nu c j$ due
  to the anti-alignment of $W$ boson spins in Higgs boson decay.
\label{fig:WW}}
\end{figure}

The key ingredient of the search for the Higgs boson in the
dilepton$+\MET$ channel is the use of the angular correlations between
the outgoing charged leptons to reduce the background coming from
direct $W^+W^-$ production.  Because the Higgs boson is a spin-0
particle, the spins of the $W$ bosons are anti-aligned.  The $V-A$
structure of $W$ boson decay causes the $W$ spin information to be
manifest in the correlated angular distributions of the decays to
leptons as seen in Fig.~\ref{fig:WW}.  Specifically, the cross section
is enhanced when the charged leptons are aligned.  There is also an
enhancement when the neutrinos align, however, the standard
dilepton$+\MET$ analysis cannot make full use of this correlation as
the neutrinos are unobserved.

The advantage of the charm-tagged semi-leptonic decay of the Higgs
boson is apparent in Fig.~\ref{fig:WW}, where we see that by
identifying the charmed jet, we may take advantage of both the
correlation between the charged lepton and the light-quark jet, and
the correlation between the charmed jet and the neutrino in the event.
Despite the fact that the neutrino appears as missing energy, we
examine cases where it comes from an on-shell $W$ decay, and can
reconstruct its four-momentum up to a two-fold ambiguity in rapidity.

\begin{figure}[htb]
\centering
\includegraphics[width=\columnwidth]{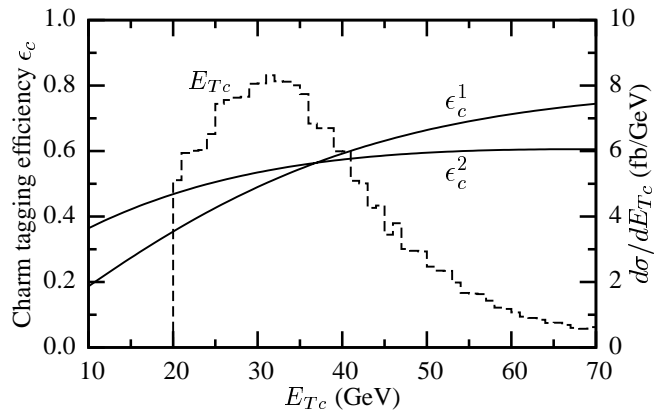}
\caption{Charm tagging efficiency curves and characteristic charm jet $E_{Tc}$
from $H\to WW\to l\nu c j$ decay. $\epsilon^1_c$($\epsilon^2_c$) are CDF
Run I(II)-like algorithms scaled up by a factor of 4.
\label{fig:ctag}}
\end{figure}

One goal of this paper is to spur development of a dedicated $c$-jet
tagging algorithm in the model of existing $b$-jet tagging algorithms.
In fact, $c$-jets are already tagged as ``fakes'' in the $b$-tagging
algorithms.  Hence, we model the transverse energy $E_T$ dependence of
the tagging efficiency utilizing existing impact-parameter $b$-tagging
algorithms from CDF Run I \cite{Carena:2000yx} (as appear in PGS 3.2
and earlier), and Run II (as appear in PGS 4) \cite{PGS4}.  Our main
results use a function of the form
\be
\epsilon^1_c =  k_c \times 0.2 \tanh \left(\frac{E_{Tj}}
{42.08 \mathrm{\ GeV}}\right) \;.
\ee
This function (shown in Fig. \ref{fig:ctag}) is a scaled version of
the PGS 3.2 impact-parameter efficiency for a $c$-jet to fake a
$b$-jet \cite{Carena:2000yx}.  In this paper we vary $k_c$ between the
Run I value of $k_c=1$, and an enhanced value of $k_c=5$ to extract
the net reach with different charm tagging efficiencies.  The
algorithm above is predominantly a fit to distributions of events in
impact-parameter vs.\ track invariant-mass \cite{wicklund}, and has
the most room for adaptation to a dedicated charm tagger.

In order to calculate the backgrounds, we utilize a $b$-tagging
efficiency of the form $\epsilon_b = k_b\times 0.6
\tanh\left(E_{Tb}/36.05 \mathrm{\ GeV}\right)$, with
$k_b=(k_c+3)/4$ in order to model saturation of $b$ acceptance.  For
light jets we take an efficiency of $\epsilon_j = 1\%\times
10^{(k_c-5)/4}$, which corresponds to a range of $0.1\%$--$1\%$.  When
$k_c=5$, the light-jet fake rate is scaled up over a factor of 10 vs.\
the baseline Tevatron rate at low jet $E_T$.  In Sec.\ \ref{sec:sim}
we will see we are completely dominated by backgrounds involving charm
jets, and hence, the details of the $b$ and light-jet backgrounds are
less important.

As a check of our main results, we have reproduced all numbers using a
CDF Run II-like algorithm $\epsilon^2_c$ from PGS 4.  The $E_T$
dependence of the charm tagging efficiency $\epsilon^2_c$ is shown in
Fig.\ \ref{fig:ctag} (scaled by a factor of 4).  Despite the different
$E_T$ dependence, we find exactly the same significances after cuts as
the backgrounds tend to have harder jets.  Since the dominance of
charm-initiated processes is insensitive to the details of the charm
tagging algorithm, we present numerical results using $\epsilon^1_c$.

An important consideration evident in Fig. \ref{fig:ctag} is that the
typical charm $E_T$ is about $20$--40 GeV.  Therefore, the effective
charm tagging efficiency used in this analysis is $\epsilon_c \sim
12$--$48\%$.  We emphasise an algorithm that improves charm tagging
\textit{acceptance} at such characteristic $E_T$'s is the relevant
efficiency and not the asymptotic value.  Neither 100\% acceptance of
$b$ jets as ``fakes'' of charms, nor a factor of 3 change to the
light-jet fake rate, materially change our results.  However, we
maintain a separate list of backgrounds so that our predictions may be
rescaled to whatever efficiencies a dedicated charm tagger eventually
obtains.

\section{Simulation and results}
\label{sec:sim}

This analysis makes use of angular correlations, and proposes the
creation of a charm tagging algorithm.  In order to accurately model
angular correlations we use the MadEvent 4.4~\cite{Alwall:2007st}
event generator.  We shower the events with PYTHIA 6.4~\cite{PYTHIA}
and use the PGS 4~\cite{PGS4} detector simulation to reconstruct
leptons and jets.  The tagging efficiencies in PGS are replaced with
the ones listed in Sec.~\ref{sec:charmtag}.  Events are generated for
$\sqrt{S}=7$~TeV and $\sqrt{S}=14$~TeV $pp$ colliders using CTEQ6L1
parton distribution functions (PDFs) \cite{Pumplin:2002vw}.  In
contrast to the dilepton studies, we allow an additional jet to be in
the event --- consistent with the effects of next-to-leading order
(NLO) radiation.

We normalize our cross sections to the NLO cross sections obtained
after acceptance cuts applied in MCFM 5.8~\cite{MCFM} using CTEQ 6.5
PDFs \cite{Tung:2006tb}.  Effective $K$-factors after cuts are shown
in Tables~\ref{tab:kfac} and \ref{tab:kfacb}.  The effective NLO
$K$-factor at 14 TeV of 1.3 for $H\to WW$ after cuts is significantly
smaller than the inclusive $K$-factor of 1.9 used by the ATLAS
Collaboration \cite{Aad:2009wy}.  Use of this smaller $K$-factor
increases our estimates of the required luminosity for $5\sigma$
discovery by a factor of 2.1 with respect to the published
experimental predictions for other Higgs decay modes at the LHC
\cite{Aad:2009wy,Ball:2007zza}.  Also note that $K$-factors for $Wbj$,
$Wcj$ and $Wc\bar c$ are estimated from jets.

\begin{table}
\caption{Next-to-leading order $K$-factors (after acceptance cuts) for
the signal and backgrounds at a 14 TeV LHC.\label{tab:kfac}}
\begin{ruledtabular}
\begin{tabular}{ccccccccc}
Signal & $Wcj$ & $WW$ & $t\bar{t}$ & $Wbj$ & t(s)-chan.& $Wc\bar{c}$ & 
$Wb\bar{b}$ & $Wjj$\\
&&&&& single top &&&\\
\hline
1.3 & 1.07 & 2.04 & 1.44 & 2.02 & 0.96(1.4) & 1.6 & 1.6 & 1.07
\end{tabular}
\end{ruledtabular}
\end{table}

\begin{table}
\caption{Next-to-leading order $K$-factors (after acceptance cuts) 
for the signal and backgrounds at a 7 TeV LHC.\label{tab:kfacb}}
\begin{ruledtabular}
\begin{tabular}{ccccccccc}
Signal & $Wcj$ & $WW$ & $t\bar{t}$ & $Wbj$ & t(s)-chan.& $Wc\bar{c}$ & 
$Wb\bar{b}$ & $Wjj$\\
&&&&& single top &&&\\
\hline
1.86 & 1.36 & 1.39 & 1.44 & 2.1 & 1.04(1.54) & 2.1 & 2.1 & 1.36
\end{tabular}
\end{ruledtabular}
\end{table}

The starting point for this analysis is the reconstruction of one
isolated lepton (an electron or muon), and two or three jets with a
single charm tag.  The angular correlations in the Higgs signal tend
to force the lepton and leading non-tagged jet to be close in phase
space.  Hence, we reconstruct jets using the PGS jet cone algorithm
with a cone size of $0.4$.  The following acceptance cuts are used to
define jets and leptons:
\be 
E_T^j > 20~\mathrm{GeV}, \; |\eta_j| < 2.5; \; p_T^l > 20~\mathrm{GeV}, \;
|\eta_l| < 2.5 \;.
\ee
Missing transverse energy $\MET$ is reconstructed from the calorimeter
and corrected for muons.

The standard model backgrounds for the semi-leptonic mode of $H \to
lcj+ \MET$ are $Wcj$, $Wbj$, $Wjj$, $Wc\bar{c}$, $Wb\bar{b}$, $W^+
W^-$, $t$- and $s$-channel single top, and $t\bar{t}$.  The
requirement of 2 or 3 jets and 1 charm tag in the final state reduces
the $t\bar{t}$ background significantly.  Charm tagging also
substantially reduces the $Wjj$ background, leaving $Wcj$ as the
dominant background at every level of cuts.  Since the Higgs decay
signal and the $Wcj$ and $Wc\bar c$ backgrounds scale with the charm
tagging efficiency $\epsilon_c$, the signal over background $S/B$ is
nearly independent of the charm tagging efficiency.  Hence, whatever
efficiency is actually obtained by the experiments will only change
the required integrated luminosity for discovery, and will not change
the analysis.  We demonstrate this point explicitly below.


Our event selection begins with a sequential set of ``common cuts''
listed in the top half of Tab.\ \ref{tab:mh165}.  We require 2 or 3
jets, with 1 charm tag and a lepton.  We require $\MET > 20$ GeV, and
reconstruct the neutrino four-momentum $p_\nu$ by fitting the lepton
and $\MET$ to an on-shell $W$ boson mass.  We take the smallest
absolute rapidity $|\eta_\nu|$ solution to complete the fit.  The
requirement that the leptonically decaying $W$ boson be on-shell
causes a significant loss of signal for Higgs bosons below $WW$
threshold, but it is necessary to reconstruct the Higgs invariant mass.

\begingroup
\squeezetable
\begin{table*}[htb]
\caption{Number of signal and background events per fb$^{-1}$ of data for
$m_H=165$~GeV and $k_c=4$ at a 14~TeV LHC, using common cuts (above line)
and ``low mass'' cuts (below line).
\label{tab:mh165}}
\begin{ruledtabular}
\begin{tabular}{l|ddddddddd}
Cuts & \textrm{Signal} & Wcj & WW & t\bar{t} & Wbj & \textrm{Single top} & Wc\bar{c} & Wb\bar{b} & Wjj\\
\hline
2 or 3 jets, 1 $c$ tag, 1 $l$ & 259  & 85192 & 2090 & 17223 & 7635 & 15166 & 
2995 & 1118 & 13331\\
$\MET > 20.0$~GeV & 228	& 74798 & 1818 & 15694 & 6862 & 13755 & 2685 & 1016 & 
12327\\
$E_{Tc} < 80$~GeV & 224 & 53563 & 1582 & 9957 & 5143 & 10103 & 2153 & 803 & 
9189\\
$p_{Tl} < 60$~GeV & 216 & 40222 & 1304 & 7074 & 4323 & 8558 & 1716 & 615 & 
6784\\
$|\Delta \eta_{cj}| < 2.0$ & 184 & 29818 & 1214 & 5921 & 3185 & 4991 & 1426 & 539 & 5012\\
$\Delta \phi_{c\nu} < 1.5$ & 142 & 11393 & 223 & 1993 & 1216 & 1470 & 393 & 
153 & 1618 \\ 
$\Delta \phi_{jl}  < 2.0$ & 126 & 8867 & 178 & 1364 & 961 & 1136 & 321 & 123 & 
1175 \\ \hline
$\cos\theta_{jc} < -0.5$ & 106 & 5178 & 78 & 846 & 488 & 740 & 160 & 70 & 659\\
$\cos\theta_{l\nu} < -0.8$ & 75 & 1875 & 42 & 181 & 170 & 145 & 73 & 31 & 198\\
$\cos \theta_l^0 < 0.2$ & 59 & 1179 & 28 & 147 & 117 & 109 & 44 & 21 & 138 \\
$140 \leq M_{l\nu cj} \leq 170$ & 45 & 423 & 19 & 6 & 39 & 18 & 20 & 9 & 58
\end{tabular}
\end{ruledtabular}
\end{table*}
\endgroup

As $Wcj$ is the most problematic background, we tune most cuts to
reduce its contribution.  In the $Wcj$ background, the transverse
energy of the charm jet $E_{Tc}$ and transverse momentum of the lepton
$p_{Tl}$ have a harder spectrum than the signal.  We therefore impose
cuts on the maximum values these variables can take: $E_{Tc} < 80$
GeV; $p_{T l} < 60$ GeV.  These values are optimal for a 160 GeV Higgs
boson, but could be loosened in a more optimized fit for larger-mass
Higgs bosons.  In the $Wcj$ background, $\Delta\eta_{cj}$, the
pseudorapidity between the charm jet $c$ and leading non-tagged jet
$j$, has a slightly broader distribution than in the Higgs signal.
Hence, we require $|\Delta \eta_{cj}| < 2$.

The backgrounds that are independent of charm tagging --- $Wb\bar{b}$,
$Wbj$, $t\bar t$, and single top --- can be reduced significantly by
using the angular correlations between the final state particles.  The
simplest angular cuts, are those similar to that of $\Delta \phi_{ll}$
in the leptonic channel, where we cut on the equivalent $\Delta
\phi_{jl} < 2$ between the lepton $l$ and \textit{leading} non-tagged
jet $j$.  In addition, we know that the directions of the neutrino and
$c$-jet are correlated.  Therefore we can also make a cut $\Delta
\phi_{c\nu} < 1.5$.  These cuts also have a strong impact on the
$Wcj$, $Wjj$, and $W^+W^-$ backgrounds.

The remaining cuts of Tab.\ \ref{tab:mh165} are specific to ``low
mass'' Higgs bosons, that is $M_H \alt 170$ GeV.  We make strong cuts
on the angle between the jet and charm $\cos \theta_{jc} < -0.5$ and
between the lepton and neutrino $\cos \theta_{l\nu} < -0.8$.  Above
170~GeV we use a set of ``high mass'' cuts shown in Tab.\
\ref{tab:mh180} that loosen the angular cuts to $\cos \theta_{jc} <
-0.2$ and $\cos \theta_{l\nu} < -0.4$, because boosts to the on-shell
$W$ bosons tend to push the peak of these distributions away from
being back-to-back.  The opening of these cuts is the main reason for
the reduction in sensitivity at larger masses.

\begingroup
\squeezetable
\begin{table*}[htb]
\caption{Number of signal and background events per fb$^{-1}$ of data for
$m_H=180$~GeV and $k_c=4$ at a 14~TeV LHC, using ``high mass'' cuts after
common cuts (top half of Tab.~\protect{\ref{tab:mh165}}) have been applied.
\label{tab:mh180}}
\begin{ruledtabular}
\begin{tabular}{l|ddddddddd}
Cuts & \textrm{Signal} & Wcj & WW & t\bar{t} & Wbj & \textrm{Single top} & Wc\bar{c} & Wb\bar{b} & Wjj\\
\hline
$\cos\theta_{jc} < -0.2$ & 81 & 6759 & 114 & 1117 & 658 & 957 & 223 & 94 & 866\\
$\cos\theta_{l\nu} < -0.4$ & 72 & 4747 & 98 & 626 & 431 & 499 & 173 & 72 & 581\\
$50 \leq M_{jc} \leq 85$~GeV & 56 & 1971 & 80 & 65 & 163 & 88 & 95 & 37 & 282\\
$m_H-20 \leq M_{l\nu cj} \leq m_H+10$ & 42 & 1167 & 50 & 40 & 93 & 53 & 52 & 20 & 163
\end{tabular}
\end{ruledtabular}
\end{table*}
\endgroup

The final angular variable we consider is the angle $\theta^0_l$ of
Ref.~\cite{Dobrescu:2009zf}.  $\theta^0_l$ is the angle between the
lepton and the boost direction of the initial $W$-boson in the
Higgs boson rest frame.  For ``low mass'' Higgs bosons we place a cut
demanding $\cos \theta_l^0 < 0.2$.  We find this angle is less
effective in the ``high mass'' regime, and therefore do not apply the
cut.  Instead, we utilize the knowledge that the second $W$ boson will
be on-shell and impose a $W$ mass reconstruction cut of $50 < M_{jc} <
85$ GeV on the search for ``high mass'' Higgs bosons.  We notice that
the peak of the reconstructed $W$ mass is below the nominal value of
$80.4$ GeV due to our use of a small cone size in our jet
reconstructions.  It may be desirable experimentally to apply jet
energy corrections and tighten this cut window, but we accept the
reduction in efficiency in this analysis.

\begin{figure}[htb]
\centering
\includegraphics[width=\columnwidth]{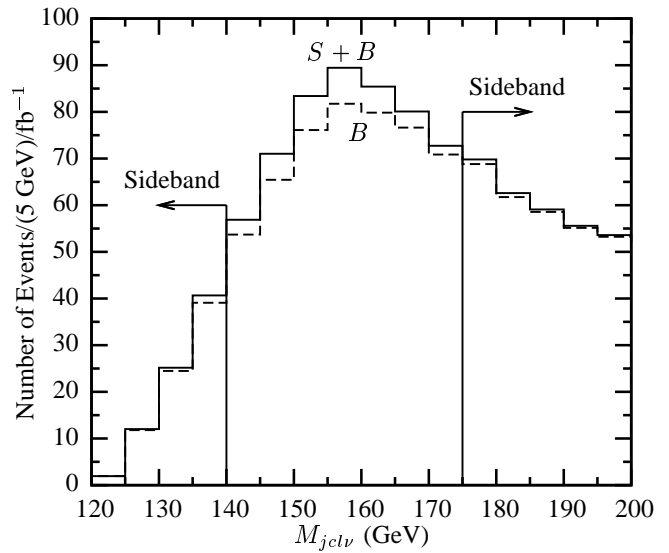}
\caption{Invariant mass of the $l\nu cj$ system for the background $B$,
and 165~GeV Higgs boson signal $S$ plus background $S+B$, for $k_c=4$.
  \label{fig:mjclnu}}
\end{figure}

We show the invariant mass distribution $M_{l\nu cj}$ of background
$B$, and a 165 GeV Higgs signal plus background $S+B$ in Fig.\
\ref{fig:mjclnu}.  We choose to leave some of the cuts looser than is
optimal in order to have a significant region of pure background on
both side-band regions of the invariant mass.  This will allow a
reasonably accurate \textit{in situ} measurement of the background,
and should greatly reduce systematic uncertainties.  We finish by
reconstructing the invariant mass of the $l\nu cj$ system, so that
$140 \leq M_{l\nu cj} \leq 170$ GeV in the ``low mass'' region, and
$m_H-20 \mathrm{\ GeV} \leq M_{l\nu cj} \leq m_H + 10$~GeV in the
``high mass'' region.

\begin{figure}[htb]
\centering
\includegraphics[width=\columnwidth]{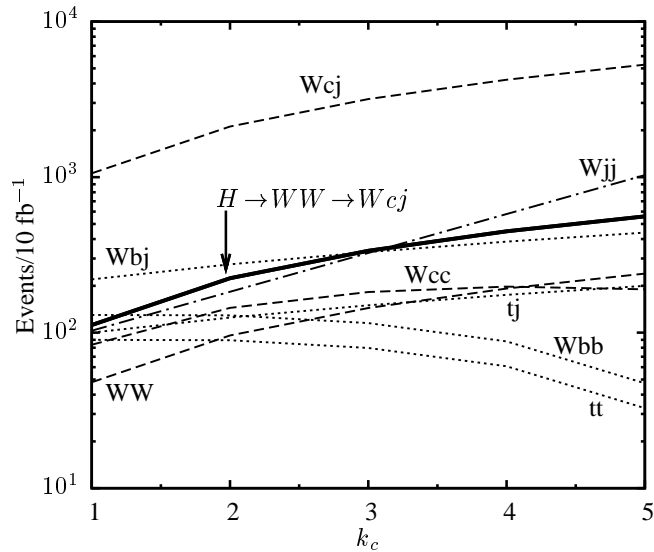}
\caption{Number of events in 10 $\fbi$ at a 14 TeV LHC for the $H\to Wcj$
signal (solid), and backgrounds from processes with charm jets (dashed),
bottom jets (dotted), and light jets (dot-dashed).
\label{fig:modes}}
\end{figure}

From Tab.~\ref{tab:mh165}, we see that the resulting signal to
background ratio $S/B\sim 1/13$ for a 165 GeV Higgs boson when the
charm tagging efficiency averages 48\% ($k_c=4$).  Before examining
the signal significance, we first demonstrate that $Wcj$ background is
an order-of-magnitude larger than any other background for all charm
tagging efficiencies.  In Fig. \ref{fig:modes} we show the expected
number of events after all cuts in 10 $\fbi$ of integrated luminosity
for the signal and all backgrounds.  With current $b$-fake rates,
$Wbj$ is slightly larger than the signal (though much smaller than
$Wcj$), but quickly saturates.  If $k_c$ is large, then $Wjj$ grows in
importance, but is also still insignificant with respect to $Wcj$.
Since the signal and direct $Wcj$ have the same charm jet tagging
efficiency, they scale together.  Hence, our observation that
luminosity required for a reaching a given significance will scale
approximately linearly with $k_c$.  All other backgrounds are small
compared to the signal for any charm jet tagging efficiency, and
backgrounds with two $b$-jets become less important as events with two
tags are discarded.

\begin{figure}[htb]
\centering
\includegraphics[width=\columnwidth]{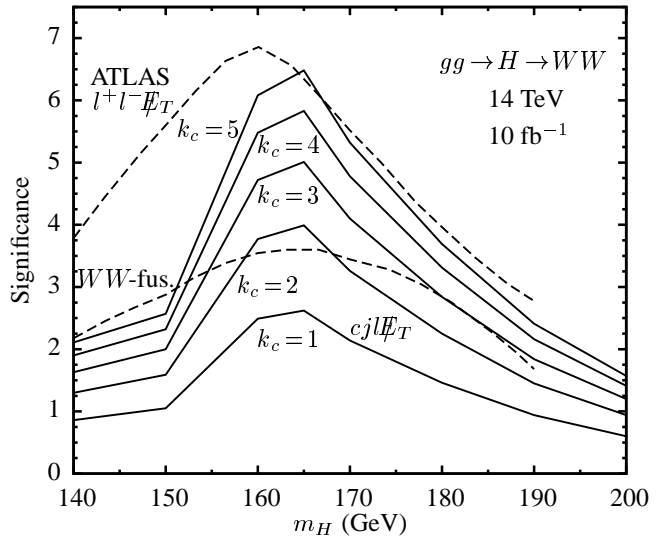}
\caption{Higgs signal significance obtainable at a 14 TeV LHC with 10 $\fbi$
of integrated luminosity vs.\ Higgs boson mass $m_H$ in the $l\nu cj$ channel
for various multiples of the current charm tagging efficiency $k_c$.  ATLAS
dilepton reach is shown (assuming a $K$-factor of $1.3$) for gluon-fusion
($l^+l^-\MET$) and $WW$-fusion \protect{\cite{Aad:2009wy}}.
\label{fig:reach}}
\end{figure}

Using the numbers in Tab.\ \ref{tab:mh165}, we find that if $k_c=4$, a
165 GeV Higgs boson could be discovered at $5\sigma$ with 7 $\fbi$ of
integrated luminosity.  In general, we are interested in what
significance can be reached as a function of Higgs mass, and what
improvement in charm tagging efficiency is required to get there.  In
Fig.\ \ref{fig:reach}, we present the significance obtainable with 10
$\fbi$ of data at a 14 TeV LHC as a function of Higgs mass.  With
current charm tagging efficiency ($k_c=1$), the $H\to Wcj$ channel
plays a complementary role in the Higgs search, contributing
$1.5$--$2.5\sigma$ to a combined analysis.  With a modest improvement
in charm acceptance ($k_c\sim 2$), corresponding to about $25\%$
acceptance, the $Wcj$ final state has comparable reach to $WW$ fusion
production with decay into $l^+l^-\MET$ for all Higgs masses.  This is
evident in Fig.\ \ref{fig:reach}, where we show the ATLAS $WW$-fusion
reach extracted from Ref.\ \cite{Aad:2009wy}, scaled to our estimate
of the NLO $K$-factor after cuts.  In addition, we show the ATLAS
gluon-fusion prediction for $H\to l^+l^-\MET$ scaled to our NLO
$K$-factor after cuts (1.3 vs.\ 1.9 used in Ref.\ \cite{Aad:2009wy}).
If charm tagging efficiency could be raised to $\sim 50$--$60\%$
($k_c=4$--$5$), then the $Wcj$ final state has comparable reach to
dileptons$+\MET$ above 155~GeV.  While improved charm tagging
efficiency is important in its own right, it is clear the $H\to Wcj$
can play a role in the Higgs search comparable to existing channels.

At present, the LHC is operating at 7~TeV, which leads to a reduction
in the cross-sections of both the signal and Standard Model
backgrounds compared to our predictions at 14~TeV.  In addition, the
lower energies also lead to a softer spectrum for the jets and
leptons.  Therefore, in Table~\ref{tab:7tevcuts} we move the charm jet
$E_T$ cut from $80$~GeV to $60$~GeV.  Once this cut is made, the
optimal lepton $p_T$ remains the same as in the 14~TeV analysis.  As
there is less energy in these events, the Higgs boson is less boosted
and so decays in a more central region of the detector.  Hence, the
cut on $|\Delta \eta_{lc}| < 1.5$, for the $7$~TeV analysis, has a
greater effect on improving $S/B$ as compared to that in $14$~TeV
analysis.

\begingroup
\squeezetable
\begin{table*}[htb]
\caption{Number of signal and background events per fb$^{-1}$ of data for
$m_H=165$~GeV and $k_c=4$ at $7$~TeV LHC.\label{tab:7tevcuts}}
\begin{ruledtabular}
\begin{tabular}{l|ddddddddd}
Cuts & \textrm{Signal} & Wcj & WW & t\bar{t} & Wbj & \textrm{Single top} & Wc\bar{c} & Wb\bar{b} & Wjj\\
\hline
2 or 3 jets, 1 $c$ tag, 1 $l$ & 32 & 3812 & 245 & 46 & 1016 & 1406 & 906 & 
345 & 932\\
$\MET > 20.0$~GeV & 30 & 3522 & 232 & 45 & 957 & 1353 & 839 & 326 & 892\\ 
$E_{Tc} < 60$~GeV & 24 & 1946 & 148 & 19 & 601 & 666 & 536 & 223 & 283\\
$p_{Tl} < 60$~GeV & 19 & 821 & 81 & 7 & 271 & 368 & 256 & 106 & 88\\ 
$|\Delta \eta_{cj}| < 1.5$ & 16 & 591 & 59 & 5 & 171 & 246 & 191 & 75 & 59\\
$\Delta \phi_{c\nu} < 1.5$ & 12 & 230 & 15 & 1.6 & 66 & 85 & 55 & 26 & 19\\
$ 2.9 < \Delta R_{jl}  < 2.1$ & 10 & 162 & 11 & 1.2 & 49 & 57 & 35 & 18
& 14 \\
$2.0 < \Delta R_{ln} < 3.0$ & 8 & 104 & 7 & 0.5 & 31 & 27 & 22 & 12 & 5\\
$m_H-20 \leq M_{l\nu cj} \leq m_H+20$ & 4.4 & 34 & 2.8 & 0.04 & 9 & 3 & 9 & 3.9 & 1.5
\end{tabular}
\end{ruledtabular}
\end{table*}
\endgroup

\begin{figure}[htb]
\centering
\includegraphics[width=\columnwidth]{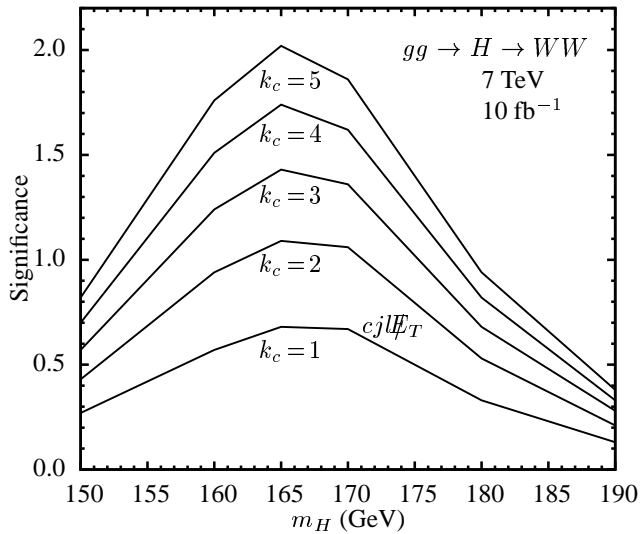}
\caption{Higgs signal significance obtainable at a 7 TeV LHC with 10 $\fbi$
of integrated luminosity vs.\ Higgs boson mass $m_H$ in the $l\nu cj$ channel
for various multiples of the current charm tagging efficiency $k_c$.
\label{fig:reachnow}}
\end{figure}

Similar to the $14$~TeV analysis, the angular correlations between
charm and the neutrino make the $\Delta \phi_{c\nu}$ cut important in
reducing the single-top and standard model $W^+W^-$ backgrounds.
However, due to the lack of energy, the remaining angular cuts in the
$14$~TeV analysis would degrade the significance of the signal.
Instead, we impose $\Delta R$ cuts shown in Table~\ref{tab:7tevcuts}
that marginally improve the signal significance.  Finally, we extract
the Higgs mass with a $40$~GeV window about the central value. Using
the numbers shown in Table~\ref{tab:7tevcuts} we see that to obtain a
95\% confidence-level exclusion limit for a $m_H = 165$~GeV we need a
$13$~fb$^{-1}$ integrated luminosity for $k_c=4$.  In Fig.\
\ref{fig:reachnow} we show the reach in the $Wcj$ final state as a
function of Higgs mass for various improvements $k_c$ to charm tagging
efficiency.  For all masses and $k_c$ the significance with 10 $\fbi$
of data ranges between $0.5$--$2\sigma$.  Hence, $Wcj$ should play a
complimentary role in the $7$~TeV LHC exclusion limits for a standard
model Higgs boson.

\section{Conclusions}
\label{sec:concl}

We demonstrate the use of a new channel for the discovery of a Higgs
boson of mass 140--200 GeV, $H\to W^+W^-\to l\nu c j$.  By utilizing a
dedicated charm tagger we gain experimental access to angular
correlations not observable in the dilepton$+\MET$ analysis that can
be used to reduce the $Wjj$ backgrounds to an acceptable level.  This
could allow a $5\sigma$ discovery of a Higgs boson near the $WW$
threshold with $\sim 7$ $\fbi$ of integrated luminosity at a 14~TeV
Large Hadron Collider in the $Wcj$ channel.  If the current 7~TeV run
of the LHC delivers 10 $\fbi$ of data, then this channel could help
rule out a standard model Higgs boson with mass between 150--190 GeV.

While one goal is to motivate the creation of a dedicated charm tagger,
this analysis is robust without any improvements to existing
algorithms.  In particular, the luminosity required for $5\sigma$
discovery scales inversely with charm tagging efficiency.  Therefore,
utilizing the charm contamination of existing $b$-tagging algorithms
as a poor charm-tagger ($\epsilon_c \sim 12\%$ for typical jet
transverse energies) already is enough to provide $1/2$ the
significance of $WW$-fusion searches, and $1/5$ the significance of
existing gluon-fusion searches to dileptons plus missing transverse
energy.  With a small improvement to charm acceptance, the $Wcj$
channel has comparable reach.  $Wcj$ provides a window into the Higgs
searches with significantly different backgrounds and low correlation
to other channels.

We conclude with the observation that the angular correlations in the
Higgs signal lead to a fairly soft missing energy signature.  In this
analysis we choose a cut of $\MET>20$ GeV, as used in the recent ATLAS
projections for large-mass Higgs reconstruction \cite{Aad:2009wy}.  We
have also investigated the effects of increasing the $\MET$ to 30~GeV,
and find the significance declines by only 5\%.  Hence, our
conclusions remain robust if higher instantaneous luminosities force
the use of harder cuts.

Given the significant correlations between the Higgs decay products,
it seems likely that more complicated multivariate techniques, such as
neural network searches, could improve the prospects for searches in
the $Wcj$ final state.  This should be pursued for integration into
the current 7 TeV Higgs searches at the LHC.  Having demonstrated the
promise of the semi-leptonic channel at the LHC, it will be useful to
see how this channel can contribute to the ongoing Higgs boson
searches at the Fermilab Tevatron.

\begin{acknowledgments}
  Z.S.\ wishes to thank the Aspen Center for Physics for support in
  developing this manuscript.  This work is supported by the U.~S.\ 
  Department of Energy under Contract No.\ DE-FG02-94ER40840.
\end{acknowledgments}


\begin{thebibliography}{99}

\bibitem{Aad:2009wy}
G.~Aad {\it et al.} (ATLAS Collaboration),
\eprint{arXiv:0901.0512}.

\bibitem{Ball:2007zza}
G.~L.~Bayatian {\it et al.} (CMS Collaboration),
J.\ Phys.\ G {\bf 34}, 995 (2007).

\bibitem{Barger:1990mn}
V.~D.~Barger, G.~Bhattacharya, T.~Han, and B.~A.~Kniehl,
Phys.\ Rev.\  D {\bf 43}, 779 (1991).

\bibitem{Dittmar:1996ss}
M.~Dittmar and H.~K.~Dreiner,
Phys.\ Rev.\  D {\bf 55}, 167 (1997).

\bibitem{Han:1998ma}
T.~Han and R.~J.~Zhang,
Phys.\ Rev.\ Lett.\  {\bf 82}, 25 (1999).


\bibitem{Sullivan:2006hb}
Zack~Sullivan and Edmond~L.~Berger,
Phys.\ Rev.\ D {\bf 74}, 033008 (2006);
Phys.\ Rev.\  D {\bf 82}, 014001 (2010).

\bibitem{Dobrescu:2009zf}
B.~A.~Dobrescu and J.~D.~Lykken,
\eprint{arXiv:0912.3543 [hep-ph]}.

\bibitem{Carena:2000yx}
M.~Carena {\it et al.} (Higgs Working Group Collaboration), 
in {\sl Physics at Run II: the Supersymmetry/Higgs Workshop}, Fermilab, 1998,
edited by M.~Carena and J.~Lykken (Fermilab, Batavia, 2002), p.\ 424.

\bibitem{PGS4}
PGS 4, \url{physics.ucdavis.edu/~conway/research/software/pgs/pgs4-general.htm}.

\bibitem{:2007dm}
T.~Aaltonen {\it et al.}  (CDF Collaboration),
Phys.\ Rev.\ Lett.\  {\bf 100}, 091803 (2008).

\bibitem{Tung:2006tb}
W.K.~Tung \textit{et al.},
J.\ High Energy Phys.\ {\bf 0702}, 053 (2007).

\bibitem{SUSYT}
V.~M.~Abazov {\it et al.}  (D0 Collaboration),
Phys.\ Lett.\  B {\bf 665}, 1 (2008);
(CDF Collaboration), CDF Note 9834, unpublished.

\bibitem{Bellazzini:2009xt}
B.~Bellazzini \textit{et al.},
Phys.\ Rev.\  D {\bf 80}, 075008 (2009).

\bibitem{wicklund}
Barry Wicklund, private communication.

\bibitem{Alwall:2007st}
J.~Alwall {\it et al.},
JHEP {\bf 0709}, 028 (2007).

\bibitem{PYTHIA}
T.~Sjostrand \textit{et al.},
Comput.\ Phys.\ Commun.\ {\bf 135}, 238 (2001);
T.~Sjostrand \textit{et al.},
\eprint{arXiv:hep-ph/0308153}.

\bibitem{Pumplin:2002vw}
J.~Pumplin \textit{et al.},
J.\ High Energy Phys.\ {\bf 0207}, 012 (2002).

\bibitem{MCFM}
J.~M.~Campbell and R.~K.~Ellis,
Phys.\ Rev.\  D {\bf 60}, 113006 (1999).



\end{thebibliography}
\end{document}